\PassOptionsToPackage{table}{xcolor}
\documentclass[sigconf]{acmart}
\makeatletter
\renewcommand{\@affiliationfont}{\large\normalfont}
\makeatother
\AtBeginDocument{%
  }

\copyrightyear{2026}
\acmYear{2026}
\setcopyright{cc}
\setcctype{by}
\acmConference[IH\&MMSec '26]{ACM Workshop on Information Hiding and Multimedia Security}{June 17--19, 2026}{Firenze, Italy}
\acmBooktitle{ACM Workshop on Information Hiding and Multimedia Security (IH\&MMSec '26), June 17--19, 2026, Firenze, Italy}
\acmDOI{10.1145/3785353.3815080}
\acmISBN{979-8-4007-2376-6/2026/06}

\DeclareMathOperator*{\argmin}{argmin}

\usepackage{booktabs}
\usepackage{enumitem}
\setlist[itemize]{leftmargin=1em,labelsep=0.5em}
\setlist[enumerate]{leftmargin=1em,labelsep=0.5em}
\definecolor{source}{HTML}{154360}
\definecolor{target}{HTML}{10B58D}
\definecolor{lighttarget}{HTML}{C7FFF1}
\definecolor{lightsource}{HTML}{CEEAFD}
\definecolor{tada}{HTML}{caffbf}

\makeatletter
\def\@mkbibcitation{\bgroup
  \let\@vspace\@vspace@orig
  \let\@vspacer\@vspacer@orig
  \def\footnotemark{}%
  \def\\{\unskip{} \ignorespaces}%
  \def\footnote{\ClassError{\@classname}{Please do not use footnotes
      inside a \string\title{} or \string\author{} command! Use
      \string\titlenote{} or \string\authornote{} instead!}}%
  \par\medskip\small\noindent{\bfseries ACM Reference Format:}\par\nobreak
  \noindent\bgroup
    \def\\{\unskip{}, \ignorespaces}\authors\egroup. \@acmYear. \@title
  \ifx\@subtitle\@empty. \else: \@subtitle. \fi
  \if@ACM@nonacm\else
    Proceedings of the 2026 ACM Workshop on Information Hiding and Multimedia Security (IH\&MMSec~'26), June 17--19, 2026, Firenze, Italy. ACM, New York, NY, USA, 6 pages.
  \fi
  \ifx\@acmDOI\@empty\else\@formatdoi{\@acmDOI}\fi
\par\egroup} 
\makeatother

\begin{document}

\title{Tackle CSM in JPEG Steganalysis with Data Adaptation}

\author{Rony Abecidan}
\email{rony.abecidan@label4.ai}
\affiliation{%
  \institution{LABEL4.AI, Univ. Lille, CNRS, Centrale Lille, UMR 9189 CRIStAL}
  \city{Lille}
  \country{France}
}

\author{Vincent Itier}
\email{vincent.itier@imt-nord-europe.fr}
\affiliation{%
  \institution{Centre for Digital System, IMT Nord Europe, UMR 9189 CRIStAL, CNRS}
  \city{Lille}
  \country{France}
}

\author{Jérémie Boulanger}
\email{jeremie.boulanger@univ-lille.fr}
\affiliation{%
  \institution{Univ. Lille, CNRS, Centrale Lille, UMR 9189 CRIStAL}
  \city{Lille}
  \country{France}
}

\author{Patrick Bas}
\email{patrick.bas@cnrs.fr}
\affiliation{%
  \institution{Univ. Lille, CNRS, Centrale Lille, UMR 9189 CRIStAL}
  \city{Lille}
  \country{France}
}

\author{Tomáš Pevný}
\email{pevnak@protonmail.ch}
\affiliation{%
  \institution{Department~of~Computers~and Engineering, CTU}
  \city{Prague}
  \country{Czech Republic}
}

\renewcommand{\shortauthors}{Abecidan et al.}

\begin{abstract}
  Steganalysis models excel on benchmark datasets but struggle in the wild when analyzed images are produced by a processing pipeline unseen during training. This problem known as Cover Source Mismatch (CSM) is particularly hard in realistic settings where practitioners  (1) have access to only a small, unlabeled dataset, (2) are unsure of the processing techniques applied to these images, and (3) lack information on the proportion of covers and stegos in that set. To answer this challenge, we introduce TADA (Target Alignment through Data Adaptation), a framework learning to emulate the unknown processing pipeline from a small unlabeled target set. This architecture is trained with a loss combining residual covariance alignment, residual distribution matching, and a $\ell^2$ loss constraining the emulator to produce realistic images. Across toy and operational targets, TADA yields substantial gains in robustness to CSM and improves operational generalization compared to strong holistic and atomistic baselines. Additional resources are available at this link: \url{https://github.com/RonyAbecidan/TADA}.
\end{abstract}

\ccsdesc[500]{Security and privacy}
\ccsdesc[500]{Computing methodologies~Image manipulation}
\ccsdesc[500]{Computing methodologies~Neural networks}

\keywords{Steganalysis, cover source mismatch, data adaptation, convolutional neural network}


\maketitle

\section{Introduction}
State-of-the-art steganalysis detectors are typically built by fine-tuning machine learning models on curated datasets like ALASKA \cite{alaska2} or BOSS\cite{boss}, where cover and stego distributions are carefully controlled. While this paradigm yields excellent benchmark performance, it overlooks a serious practical constraint: Operational sets encountered in the wild display strong heterogeneity driven by acquisition devices, sensor quality, camera settings, scene characteristics, editing operations, and compression workflows. Each step modifies the underlying statistics of the covers and impacts steganographic artifacts. This situation leads to the Cover-Source Mismatch phenomenon \cite{csmmallet}: detectors trained on a specific cover-stego distribution often struggle to generalize when faced with a different distribution, a widely acknowledged challenge in machine learning known as out-of-distribution generalization.

\begin{figure*}[t]
  \centering
  \includegraphics[width=0.8\textwidth,height=0.75\textheight,keepaspectratio]{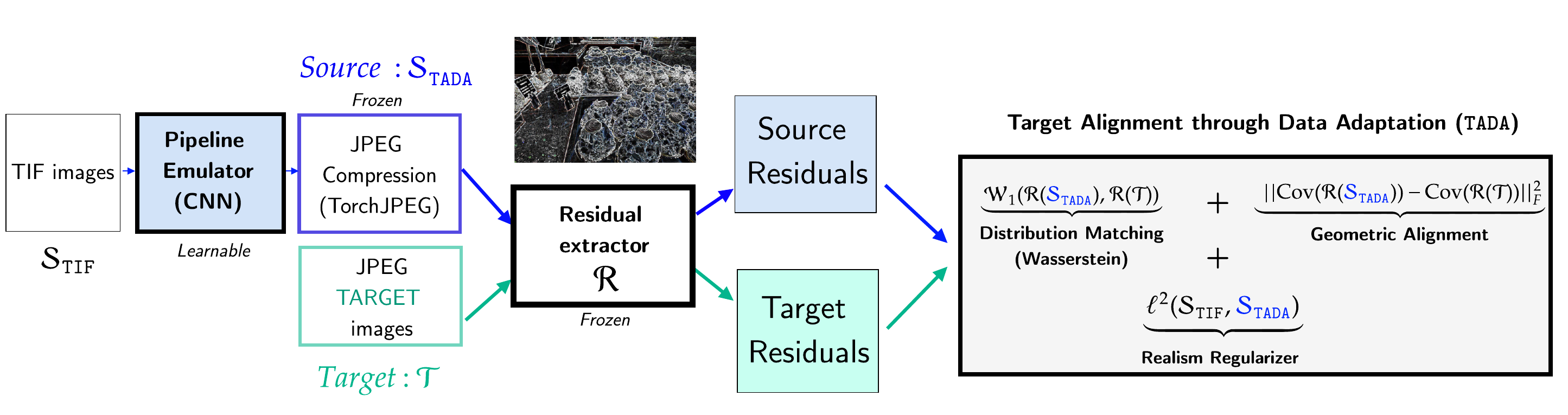}
  \caption{\texttt{TADA} learns a lightweight convolutional emulator so that residual statistics of the emulated source match those of the target, thereby reducing CSM.}
  \label{fig:tada}
\end{figure*}

\subsection{Prior Methodologies to fight CSM}

To fight CSM, the literature mainly distinguishes two families: \emph{holistic} and \emph{atomistic} strategies. Holistic methods aim to improve robustness by training on mixtures of sources that increase content/noise diversity; ALASKA~\cite{alaska2} is a representative example with 80k images captured with 479 sensors. Yet, diversity alone is not sufficient:~\cite{wifs2022} shows that some combinations are ineffective, while carefully selected small mixtures can generalize better in the wild. In practice, however, choosing the right mixture for a specific operational target is difficult when little information is available about the images under scrutiny.


 
Atomistic methods instead aim to estimate, for each testing base (or even each image in the testing base), which \emph{training base distribution} is the closest, so that the steganalyst can train or deploy a detector on a more relevant basis. This typically amounts to selecting a training base from a predefined pool. For instance, Giboulot \emph{et al.}~\cite{giboulotcsm} train a multilinear classifier that assigns each test image to one of several representative training bases and then applies the detector trained on that representative. Similarly, Abecidan~et~al.~\cite{wifs2023} guide training base selection via simulated annealing using the chordal distance between DCTr features~\cite{dctr} of candidate training bases and the testing base. A key limitation is that atomistic selection hinges on accurate prior knowledge on the testing base processing pipeline (operations and hyperparameters). This requirement is often unrealistic in the wild, and~\cite{giboulotcsm} notably shows that even a slight error in JPEG quality factor can substantially degrade generalization.

Overall, both families can fail when the available training bases (or their mixtures) remain mismatched with the operational testing base, motivating the need to create a specific training basis rather than selecting one from a predefined pool.


\subsection{Contributions}

While several strategies are proposed to address 
CSM in steganalysis \cite{csmmallet}, few are designed for realistic operational scenarios where (i) the processing pipeline of the images under scrutiny is unknown, (ii) the analyst only has access to a small operational set produced by that pipeline, and (iii) the labels are unavailable and the cover/stego balance may be highly skewed. Faced with that reality, our contributions are twofold:
\begin{enumerate}
  \item We introduce \texttt{TADA} (Target Alignment through Data Adaptation), an unsupervised data-adaptation strategy that turns a generic RAW image pool into a target-tailored source for JPEG steganalysis.
  \item We present a differentiable and label-free training loss that correlates with operational generalization and remains robust to unknown cover/stego balance. It combines \emph{geometric} alignment and \emph{distribution} matching on noise residuals with a realism regularizer that discourages degenerate pipelines from producing overly saturated developed images.
\end{enumerate}
\noindent
As far as we know, this data adaptation strategy is the first effort to address CSM by proposing a neural architecture designed to emulate a relevant training database with desired target statistics, especially in cases where our knowledge about these targets is very limited. Toy and real-world experiments underscore the potential of \texttt{TADA} throughout our experiments in section \ref{sec:experiments}.



\section{Context}
\subsection{Formalization}

Traditional processing of raw images can be divided into 3 consecutive steps that can be seen as a sequence of successive parameterized operations : demosaicking, aesthetics operations and JPEG compression. We propose to encapsulate the successive hyperparameters of each operation into a global vector $\boldsymbol{\omega}$ as suggested by~\cite{csmforma}. In the context of steganalysis, we also introduce  $\gamma$ to represent steganographer choices, including the embedding strategy and payload. 
We call \emph{source} ($\mathcal{S}$) the training set used to learn a detector, and \emph{target} ($\mathcal{T}$) the operational set on which this detector is evaluated. We also assume that all images in the target set are produced by the same processing pipeline $\omega_t$ and, if stego, embedded with the same strategy $\gamma_t$. In other words, the target set $\mathcal{T}(\omega_t,\gamma_t)$ is homogeneous.
For a given source, distinguishing between covers and stegos is usually done using machine learning models acting as detectors:
\[
\begin{aligned}
    f(x \mid \theta_{\omega,\gamma}) :\ & \mathcal{X} \rightarrow \{cover,stego\}, \\
    & x \mapsto y.
\end{aligned}
\]
Here, $\theta_{\omega,\gamma} \in \Theta$ represents the learned parameters using cover and stegos from a source with pipeline $\omega$ and embedding strategy $\gamma$.
\\
To effectively evaluate the CSM, two significant metrics have been introduced in \cite{giboulotcsm} and \cite{csmforma}:

\begin{itemize}
    \item The Intrinsic Difficulty of a source with pipeline $\omega$ and embedding strategy $\gamma$, representing the probability of error $P_E$ obtained after training on images from this source and evaluating on images from that same source:
    $$\mathbb E_{(x,y) \sim P((x,y)| \omega,\gamma)} (f(x \mid \theta_{\omega,\gamma} ) \neq y).$$
    \item The Regret  $\mathcal R(\mathcal{S},\mathcal{T})$ between a source set $\mathcal{S}(\omega_s,\gamma_s)$ and a target set $\mathcal{T}(\omega_t,\gamma_t)$, denoted as $\mathcal R(\mathcal{S},\mathcal{T}) \geq 0$, defined as the difference between the $P_E$ we obtain by 
    training on the source $\mathcal{S}(\omega_s,\gamma_s)$ and evaluating on the target $\mathcal{T}(\omega_t,\gamma_t)$ and the Intrinsic Difficulty of $\mathcal{T}(\omega_t,\gamma_t)$:
    $$\mathbb E_{(x,y) \sim P((x,y)|\mathcal{T})} (f(x \mid \theta_{\mathcal{S}} ) \neq y) -\mathbb E_{(x,y) \sim P((x,y)| \mathcal{T})} (f(x \mid \theta_{\mathcal{T}} ) \neq y).$$
    
\end{itemize}
In this study we want to find a source $\mathcal{S}^*(\omega_s,\gamma_s)$ minimizing the regret w.r.t. a given target $\mathcal{T}(\omega_t,\gamma_t)$. 
\begin{equation}
\mathcal{S}^*(\omega_s,\gamma_s) = \argmin_{\mathcal{S}} \mathcal R(\mathcal{S},\mathcal{T})
\end{equation}
To keep the setting realistic yet tractable, we make three assumptions. First, target images are grayscale, making the steganographer less detectable since fewer channels could betray him. Second, target images in a given operational set share the same JPEG quantization table; this is straightforward to verify in practice because quantization tables are directly readable from JPEG bitstreams. Finally, following Kerckhoffs' principle, the analyst knows the embedding strategy and parameter $\gamma_t$, so any observed Cover Source Mismatch (CSM) is entirely due to processing-pipeline mismatch. Under these assumptions, our objective is to estimate the target processing pipeline $\omega_t$ as accurately as possible. Concretely, we design a neural architecture trained to develop TIF images into possible realization of decompressed JPEG images in a relevant way, producing a source that minimizes regret with respect to the given JPEG target.

\section{Data Adaptation Framework}

\begin{figure}[h!]
  \centering
  \includegraphics[width=\linewidth]{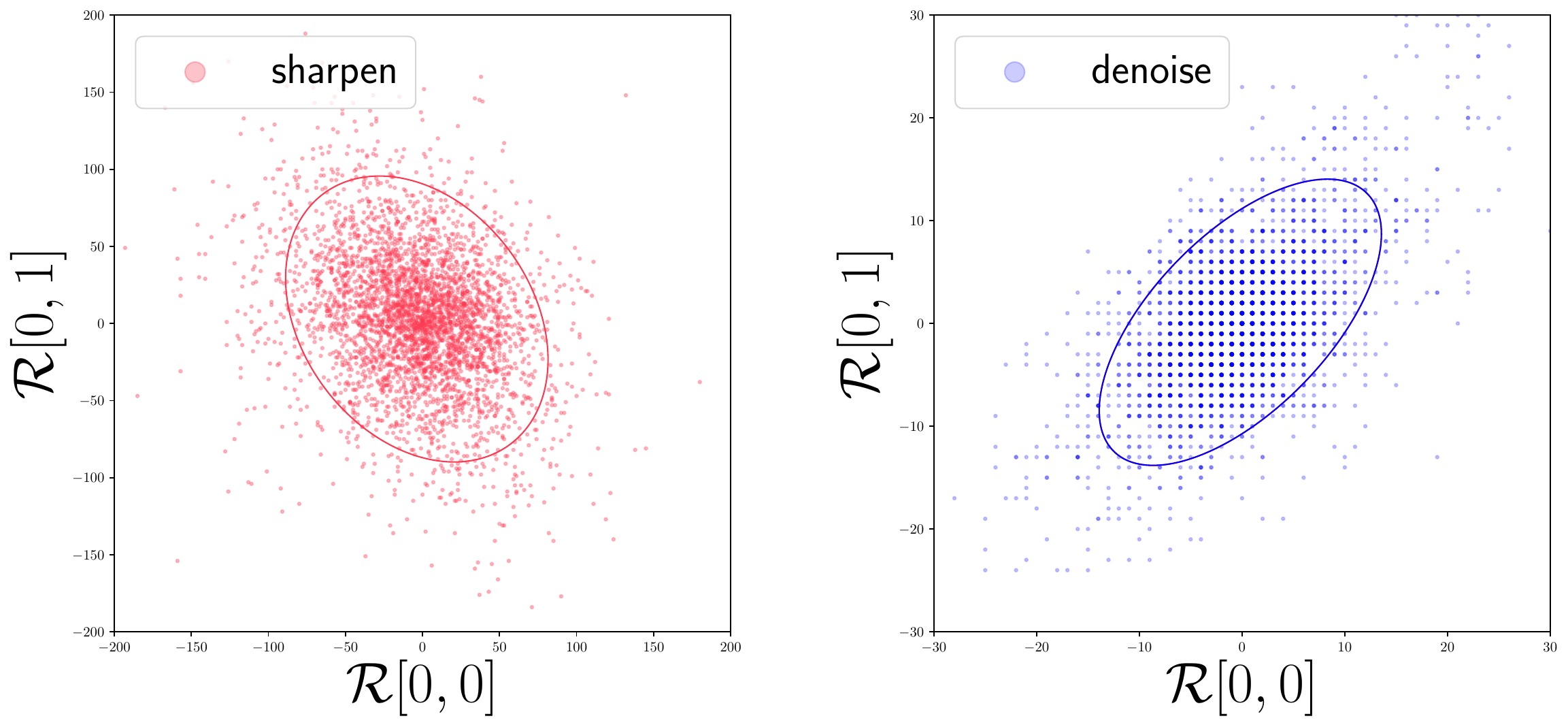}
  \caption{2D scatter plots of neighboring-pixel residuals (Laplacian $\mathcal L_4$~\cite{malletcorrelation}) within $8\times 8$ JPEG blocks for two images from the same RAW but processed differently (denoising vs.\ sharpening). Denoising induces positive residual correlations, while sharpening yields negative ones.}
  \label{fig:noise_scatter}
\end{figure}
\noindent
In this section, we describe the key components of our data adaptation architecture, \texttt{TADA}. 

\subsection{TADA Architecture}
Previous research in steganalysis showed that demosaicking operations are not the primary factor responsible for CSM~\cite{giboulotcsm,wifs2022}. Therefore, we propose to use the popular \texttt{amaze} demosaicking algorithm~\cite{amaze} to produce TIF images from RAW files as our starting point for the pipeline emulator. Afterwards, we design an architecture dedicated to TIF image development, yielding a source $\mathcal{S}_{\mathtt{TADA}}$  optimized to minimize regret against a predefined target set $\mathcal{T}$. 

The architecture requires three key components: a pipeline emulator, a pipeline fingerprint extractor, and a loss function between pipeline fingerprints that correlates with $\mathcal R(\mathcal{S}_{\mathtt{TADA}},\mathcal{T})$. The overall architecture is illustrated in Figure~\ref{fig:tada}. The data flow is decomposed into the following stages:

\begin{enumerate}
    \item First, a set of colored TIF images are processed through the \textnormal{pipeline emulator}. A differential JPEG compression from~\cite{diffjpeg} immediately follows, using the quantization table used for target images. \textnormal{At the end of the learning, the frozen pipeline emulator combined with the JPEG compression constitute our tailored processing pipeline for the target}.
    \item Afterwards, both source and target images are processed through a \textnormal{noise residual extractor}, generating characteristic features of their respective processing pipelines while being invariant to embedding.
    \item Source and target residuals are then used to compute a \textnormal{triplet loss} function that serves three key purposes: firstly, it fosters the network to progressively align geometrically source and target residuals, secondly, it brings the distributions of source and target residuals closer together and thirdly, it ensures the produced images are realistically developed, forcing the network to generate natural and plausible outputs.
\end{enumerate}
\noindent
The following subsections detail and justify our choice for each of these components.


\subsection{Choice of the pipeline emulator}

Common image processing operations (e.g., blurring, denoising, sharpening, \textit{etc.}) are differentiable operations between pixels that we can replicate with convolutions. Although these operations are not always linear, there exist linear versions of them such as mean filtering or unsharp masking, largely used in classical software like Photoshop, GIMP, and RawTherapee. More precisely, \textnormal{the bulk of processing operations can be reasonably approximated with symmetric convolutions summing to 1 as assumed in~\cite{faridthumbnail}}. For instance, $3 \times 3$ kernels satisfying these conditions are structured as $\scriptstyle\bigl[b,\,c,\,b;\ c,\,a,\,c;\ b,\,c,\,b\bigr]$ with $a+4(b+c)=1$. Considering that successive linear operations can be condensed in one linear operation, we propose using a unique convolution of this shape for the pipeline emulator of $\mathtt{TADA}$. \textit{This choice is very simple, but the reader will see 
in the next section that this constrained convolution is already very effective.}


\subsection{Choice of pipeline fingerprint extractor}
To learn a relevant target pipeline, \texttt{TADA} needs a \emph{fingerprint} that (i) is sensitive to the development pipeline, yet (ii) remains robust when the target set contains stegos. We propose here to build this fingerprint from \emph{noise residuals}.

Immediately after image acquisition, noise at the pixel level is independently distributed and follows a Poisson/Gaussian distribution~\cite{noiseraw}. Once a processing pipeline is applied to these RAW images, this initially independent noise becomes \emph{correlated} through inter-pixel operations, as illustrated in Figure~\ref{fig:noise_scatter}.

In steganalysis, it is known that noise residuals are highly sensitive to development pipelines and exhibit characteristic correlations that are robust to embedding; this is experimentally verified in~\cite{malletcorrelation,taburet}. Intuitively, many embedding schemes are designed to remain statistically close to the cover distribution: even early methods such as Model-Based steganography~\cite{sallee2003model} or nsF5~\cite{fridrich2007statistically} aim to preserve at least second-order moments of the cover distribution, or even the whole marginal distribution. As a result, residual correlations induced by the pipeline provide a usable fingerprint even when the target set contains stegos.

A simple way to extract such residuals is to apply a high-pass filter to suppress image content and emphasize inter-pixel noise dependencies. In particular, the \textnormal{KB filter}~\cite{kbfilter} is a well-known high-pass filter designed to remove image content while highlighting inter-pixel correlations, with kernel
$\mathcal{E}_{\mathrm{KB}}=\tfrac{1}{4}\scriptstyle\bigl[-1,\,2,\,-1;\ 2,\,-4,\,2;\ -1,\,2,\,-1\bigr]$. In what follows, we denote this residual extractor by $\mathcal{E}$ and use it to attenuate embedding traces while highlighting the noise correlations induced by processing pipelines.

\subsection{Choice of \texttt{TADA} loss function}
\label{sec:tadaloss}
\textnormal{Since different processing pipelines affect noise correlations differently, Mallet \emph{et al.}~\cite{malletcorrelation} proposed to treat these correlations as fingerprints of processing pipelines}. This motivates a natural first objective, namely matching residual correlations through the differentiable cost $||\operatorname{Corr}(\mathcal{E}(\mathcal{S}))-\operatorname{Corr}(\mathcal{E}(\mathcal{T}))||_{F}$. However, correlation matching is inherently ambiguous: correlations are invariant to global shifting and scaling, so two residual sets can share identical correlations while still differing substantially in magnitude. In practice, this can strongly misguide \texttt{TADA} towards pipelines that match correlation patterns but remain mismatched in residual energy.

We resolve this limitation by combining (i) \emph{covariance} alignment and (ii) \emph{distribution} matching between residuals. Covariances are shift-invariant but, unlike correlations, they are not scale-invariant: they encode both the geometry (eigenvectors) and the variance amplitudes (eigenvalues) of residuals. Distributional distances are, in turn, sensitive to shifts in residual statistics. Together, these two terms are complementary, embedding-invariant, and strongly predictive of regret in the \texttt{DCTr} space, as demonstrated in~\cite{wifs2023}. Additional elements support this choice for our training loss.

First, Giboulot \emph{et al.} showed in~\cite{giboulotnormal} that \textnormal{processed noise can be modelled with a multivariate Gaussian distribution assuming that the processing pipeline to estimate is both linear and stationary}. With \texttt{TADA}, we follow this hypothesis of linearity and stationarity by designing our emulator as a simple convolution meeting these two constraints. Assuming target residuals also follow a normal distribution, it makes sense to match source and target residual means and covariances. Second, Abecidan \emph{et al.} \cite{wifs2023} showed that the chordal distance between PCA subspaces of DCTr features \cite{dctr}, computed from source and target sets is strongly correlated with operational regret across a wide variety of sources. Although this metric is not easily differentiable, PCA projections are intrinsically governed by the covariance structure of the data. Consequently, aligning source and target covariances implicitly minimizes the chordal distance between their respective PCA subspaces. In addition, the two alignment terms alone can drive the emulator towards unrealistic pipelines, producing images that saturate after JPEG compression. To keep generated images within the natural 16-bit TIF range (no negative values and no saturation beyond $2^{16}-1$), we thus add a regularization term: an $\ell^2$ penalty between normalized TIF images and their normalized JPEG-developed counterparts. Hence, the $\mathtt{TADA}$ training loss is:
\begin{align*}
\small
    \mathcal{L} &= \lambda
    \underbrace{\|\operatorname{Cov}(\mathcal{E}(\mathcal{S})) - \operatorname{Cov}(\mathcal{E}(\mathcal{T}))\|_F^2}_{\text{\small Geometric alignment}} 
    \ + \ \mu\underbrace{d(\mathcal{E}(\mathcal{S}), \mathcal{E}(\mathcal{T}))}_{\substack{\text{\small Distribution alignment} \\ \text{(e.g., $\mathrm{MMD}$, Wasserstein)}}} \\ &+ \ \gamma  \underbrace{\ell^2(\mathcal{S}_{\mathtt{TIF}},\mathcal{S}_{\mathtt{TADA}})}_{\text{to promote output realism}}
    \text{with } \mu,\lambda,\gamma \text{ to tune}.
\end{align*}

\section{\texttt{TADA} in practice}
\label{sec:experiments}

To validate the potential of $\mathtt{TADA}$, we propose now to build \textnormal{toy and real-world targets} for which we would like to craft tailored sources.

\subsection{Evaluation process}
From each target, we derive benchmark training/evaluation splits and an unlabeled operational subset used to train $\mathtt{TADA}$. Sources are built by developing 2,000 ALASKA\cite{alaska2} RAWs into $512\times512$ smart crops using the procedure of~\cite{alaska2}. For each target we use 1,000 cover--stego pairs for benchmark training, 500 pairs for evaluation, and 500 unlabeled operational images; we consider three operational balances (all covers / all stegos / balanced mix). To contextualize \texttt{TADA}'s performance, we compare it against a naive baseline that represents the simplest possible source construction: we directly compress TIF images from ALASKA\cite{alaska2} using the target's JPEG quantization table. This baseline ignores all processing steps between RAW acquisition and JPEG compression assuming that matching only the quantization table is sufficient. 

Throughout all experiments, we use UERD~\cite{uerd} with a payload of 1 bit per non-zero AC DCT coefficient (\texttt{bpnzac}). This payload is particularly challenging for pipeline estimation since embedding artifacts can mask the subtle noise correlations that reveal processing pipelines, yet it remains realistic in operational settings where steganographers exploit CSM to their advantage: mismatched detectors become less sensitive to embedding traces, allowing higher payloads to go undetected. 

\subsection{Training details}

\textnormal{To estimate residual statistics reliably, we randomly select 500 RAWs from ALASKA~\cite{alaska2}, demosaic them with \texttt{amaze}, and extract a $512\times512$ crop as uniform as possible. These TIFs form a common base that we develop with candidate pipelines. Since orthogonal rotations preserve pipeline effects, we augment both source and target by four rotations during training.}

To avoid saturation, we initialize the emulator with the identity filter and enforce kernel constraints at the end of each epoch. Developed TIFs are then JPEG-compressed using differentiable JPEG~\cite{diffjpeg} with the target quantization table.
After JPEG compression, we compute residuals with the KB filter~\cite{kbfilter} and estimate covariances from $8\times16$ patches aligned on the JPEG grid to capture intra-/inter-block dependencies, as recommended in~\cite{taburet}.
Among the extracted patches, the lowest-variance ones typically come from highly uniform regions and bring little discriminative power to separate emulated pipelines: since our kernels have weights summing to one, residuals in uniform areas are near-null regardless of the underlying pipeline. Conversely, high-variance patches are usually textured and yield more pipeline information, but their residuals are less stable under steganographic embedding. Patch selection for covariance estimation therefore trades off pipeline discriminability and embedding robustness. To navigate this trade-off, we leverage probability maps that we can compute on target images simulating the UERD scheme used by the steganographer (assumed known under Kerckhoffs' principle) and retain only patches whose pixel standard deviation exceeds 1 and whose embedding probabilities remain below 0.01; the resulting set is both representative of the target distribution and robust to embedding. 

The training loss is computed on the selected patches.
For distribution matching we use the Earth Mover distance ($\mathcal W_1$) to avoid vanishing/exploding gradients observed with other differentiable distances~\cite[Figure 3.11]{feydy}. 
Finally, we normalize each term by its first-batch value to put all costs on comparable scales and reduce tuning effort for $\lambda, \mu, \gamma$. We train with mini-batches of 256, SGD (lr $10^{-3}$), a fixed seed (\texttt{2026}), and up to 3000 epochs, with early stopping when the monitored metric does not improve for 200 epochs.
\subsection{Toy experiments}
We construct toy targets by applying 3$\times$3 convolution kernels to TIF images before JPEG compression (quality factor 100). Two kernels are considered: one simulating denoising and another simulating sharpening. These kernels were selected to create significant mismatch with a naive baseline that simply compresses TIF images at QF100.
\noindent
To evaluate $\mathtt{TADA}$ under challenging conditions, we consider a worst-case scenario where all target images are stegos. This setup demonstrates that $\mathtt{TADA}$ can still converge to an accurate source estimate even when target samples contain strong steganographic embedding. Target kernels and $\mathtt{TADA}$ kernels are presented in Table~\ref{tab:toy} as well as the regrets obtained on these targets using the naive source and the source obtained with $\mathtt{TADA}$. 

\begin{table}[h]
\centering
\caption{Toy targets: original vs.\ $\mathtt{TADA}$-learned $3\times3$ kernels (denoising, sharpening) and the resulting target regret (\%) for naive vs.\ $\mathtt{TADA}$ sources.}
\small
\setlength{\tabcolsep}{3pt}
\renewcommand{\arraystretch}{0.95}
\begin{tabular}{|c|c|c|}
\hline
 & \multicolumn{2}{c|}{\textbf{Operations}} \\
\cline{2-3}
 & \textbf{Denoising} & \textbf{Sharpen} \\
\hline
\textbf{Original Kernel} &\resizebox{2.1cm}{!}{$\begin{bmatrix}
  0.0625 & 0.125 & 0.0625 \\
  0.125 & 0.25 & 0.125 \\
  0.0625 & 0.125 & 0.0625
  \end{bmatrix}$} & \resizebox{2.1cm}{!}{$\begin{bmatrix}
  0 & -0.25 & 0 \\
  -0.25 & 2 & -0.25 \\
  0 & -0.25 & 0
  \end{bmatrix}$} \\ 
\hline
\textbf{\texttt{TADA} Kernel} & \resizebox{2.1cm}{!}{$\begin{bmatrix}
  0.042 & 0.105 & 0.042 \\
  0.105 & 0.41 & 0.105 \\
  0.042 & 0.105 & 0.042
  \end{bmatrix}$} & \resizebox{2.1cm}{!}{$\begin{bmatrix}
  0.054 & -0.37 & 0.054 \\
  -0.37 & 2.26 & -0.37 \\
  0.054 & -0.37 & 0.054
  \end{bmatrix}$} \\
\hline
\textbf{Regret Source Naive} & 50 & 27 \\
\hline
\textbf{Regret Source TADA} & 2 & 0 \\
\hline
    \end{tabular}
    \label{tab:toy}
    \end{table}

Table~\ref{tab:toy} shows that $\mathtt{TADA}$ recovers kernels that preserve the \emph{structure} of the true operations despite strong embedding in the target. For denoising, it learns a smoothing pattern (positive weights with near-zero corners and a center coefficient $<1$); for sharpening, it recovers the key contrast pattern (strong positive center, negative neighbors; 2.26 vs.\ 2.0 in the original). This is sufficient to reduce regret to 2\% and 0\% (vs.\ 50\% and 27\% for the naive source).

\subsection{Operational experiments}

\begin{table}[h]
  \caption{Details about Flickr targets (YFCC100M).}
  \centering
  \begin{tabular*}{\columnwidth}{@{\extracolsep{\fill}}ccc}
    \toprule
      \textbf{Target name} & \textbf{Camera Model} & \textbf{Quality Factor} \\
    \midrule
      SONY & SONY SLT A37 & 90  \\
      NIKON & NIKON D40 & 90 \\
      CANON & Canon PowerShot SX30 IS & 93   \\
  \bottomrule
  \end{tabular*}
  \label{tab:rw}
\end{table}
\noindent
Real-world targets are built using the database YFCC100M \cite{yfcc100m} gathering millions of Flickr images under CC licenses. From this database, we look for users sharing non-resized images with public quantization tables to comply with our assumptions. This scenario complies with the reality of practitioners since the aesthetics operations used by Flickr users are totally unknown to us. \textnormal{We finally found 3 users sharing thousands of pictures with the same camera model and compressed with the same quantization tables}. Details about these operational targets are presented in Table~\ref{tab:rw}. All target images are initially RGB, and we convert them to grayscale to make the steganalysis task more challenging. Since the grayscale conversion applied by the steganographer is unknown, we also design \texttt{TADA} to learn a convex combination of the three color channels that best replicates this processing.
\noindent
We assume here access to 500 unlabeled $512\times512$ target images. To select the convolution size, we compared $3\times3$, $5\times5$, and $7\times7$ kernels under the same training protocol. Since the $5\times5$ setting yielded the minimum training loss, we adopted it in the operational experiments.
We compare the resulting target regret to state-of-the-art CSM baselines under operational sets that may be highly unbalanced.

\subsection{\texttt{TADA} vs SOTA strategies}

We benchmark $\mathtt{TADA}$ against state-of-the-art holistic and atomistic strategies. The key question is the same for all baselines: \emph{given a target, how do we build or pick the most relevant source?}

\begin{table}[h!]
    \centering
    \small
    \setlength{\tabcolsep}{4pt}
    \resizebox{\linewidth}{!}{%
    \begin{tabular}{l|l|l|l}
    \toprule
       & {SONY}& {NIKON} & {CANON}\\
    \midrule
        \rowcolor{lighttarget} {Intrinsic Difficulty} & 1 & 3 & 1 \\
        \rowcolor{lightsource} {Regret naive source} & 37 & 30 & 45 \\
        \midrule
        Regret $\min\limits \ \mathrm{Chordal}$ source \cite{wifs2023} & 27 & 17 & 48 \\
        {Regret All source}  \cite{wifs2022} & 14 & 7 &  36 \\
        {Regret Multiclassifier source} \cite{giboulotcsm} & 27 & \textbf{0} & 35 \\
        \midrule
        \rowcolor{tada} Regret $\mathtt{TADA}$ source (Ours, full cover) & \textbf{7} & 9 & 27  \\
        \rowcolor{tada} Regret $\mathtt{TADA}$ source (Ours, full stego) & 9 & 6 & \textbf{24}  \\
        \rowcolor{tada} Regret $\mathtt{TADA}$ source (Ours, mix) & 8 & 7 & 26  \\
    \bottomrule
  \end{tabular}%
    }
    \caption[Comparison of $\mathtt{TADA}$ vs.\ CSM baselines.]{Target regrets (\%) for CSM baselines and $\mathtt{TADA}$ across three operational regimes (full cover / mix / full stego). Best per target in bold.}
    \label{large-table}
    \end{table}
\noindent \\
\textit{\small Holistic baseline (\textbf{All}).} We start from the universe of 1,000 sources introduced in~\cite{wifs2023}. Following the greedy selection procedure of~\cite{wifs2022}, we extract 8 representative sources that are expected to foster broad generalization. We then build the \textbf{All} baseline by mixing these 8 representatives into a single, diverse training set.\\
\textit{\small Atomistic baselines (per-target source selection).} Atomistic strategies operate at the target level: they select \emph{one} representative source for each target sample (or target set), and then apply the detector trained on that representative. Using the same pool of 8 representatives, we compare two state-of-the-art selection mechanisms. First, the multiclassifier strategy of~\cite{giboulotcsm} trains an assignment model that maps each target image to one of the 8 sources based on its content; the final prediction for that target image is then produced by the detector trained on the assigned representative. Second, the chordal-distance strategy of~\cite{wifs2023} selects the representative that minimizes the chordal distance between DCTr features extracted from the target and from each representative, and uses the corresponding detector for all target images. Both approaches have shown strong performance in~\cite{wifs2023}, making them competitive baselines for $\mathtt{TADA}$. \\
\textit{\small Common evaluation protocol (fair comparison).} Across all experiments in this section, we fix the training set size to 1,000 cover-stego pairs for every method, and we derive all training bases from the same underlying RAW image set. Under our JPEG assumption, we also match the JPEG quantization tables of source images to those of target images. For $\mathtt{TADA}$, we first estimate the source pipeline parameters from the unlabeled operational target set, then generate the corresponding adapted source, and only then train the steganalysis detector on that source. For the \textbf{All} baseline specifically, the 1,000 pairs are obtained by sampling 125 pairs from each of the 8 representatives. This protocol prevents artificial gains due to larger training sets and isolates the effect of pipeline selection. Finally, all detectors are logistic regression classifiers trained on DCTr features.

Holistic and atomistic strategies used for this experiment are embedding-invariant by design: they select or construct sources independently of the cover-stego balance in the operational set, since their selection mechanisms (greedy selection, multiclassifier assignment, or chordal distance) operate on image features that are robust to embedding. 
In contrast, $\mathtt{TADA}$ learns the processing pipeline from the operational set itself, which may contain covers, stegos, or both. We thus evaluate $\mathtt{TADA}$ under all three scenarios (full cover, full stego, and balanced mix) to assess its robustness to this uncertainty. Table~\ref{large-table} reflects this difference: holistic and atomistic methods appear once, while $\mathtt{TADA}$ results are shown for all three operational set compositions.
\\
\noindent
\textit{\small Interpretation of results.} Table~\ref{large-table} yields three takeaways. \\ (i) The naive source construction confirms that matching only the JPEG quantization table is not enough to mitigate CSM (30--45\% regret). (ii) Holistic/atomistic baselines are highly target-dependent: \textbf{All} works well on Nikon (7\%) and Sony (14\%) but not on Canon (36\%), while atomistic selection can be excellent (0\% on Nikon) yet fail on other targets when the true pipeline is not well represented in the source pool. (iii) $\mathtt{TADA}$ is the most consistent across targets and unknown cover/stego balance, reaching single-digit regrets on Sony/Nikon (6--9\%) for all three operational regimes and the best Canon regrets (24--27\%), though Canon remains harder to emulate with our current lightweight emulator.

\subsection{Limitations and Perspectives}

While the results above are encouraging, our current implementation makes several simplifying choices that outline clear directions for future work. Both the $\mathtt{TADA}$ emulator and the residual extractor are deliberately simple, which supports stable optimization and interpretability but limits the pipelines we can capture (e.g., resizing or strongly non-linear operations). Moreover, $\mathtt{TADA}$ assumes a homogeneous operational set; if the target mixes several development pipelines, a practical perspective is to first cluster images by pipeline fingerprints and then run $\mathtt{TADA}$ per cluster. In operational experiments, $\mathtt{TADA}$ did not reliably converge to a relevant pipeline in the \textit{mix} and \textit{full stego} settings without our patch selection criterion. Selecting patches that are both pipeline-discriminative and relatively robust to embedding is thus critical, and improving (or learning) this selection remains an important perspective. Reducing the number of operational patches degrades the learned pipeline, although simple augmentation keeps performance acceptable with roughly half as many patches. The final loss was also chosen for practical robustness: using only Wasserstein or only geometric alignment is weaker, while the realism term remains important to avoid less plausible or saturated emulated images.

\section{Conclusion}

We presented $\mathtt{TADA}$, a data-adaptation framework that mitigates CSM by learning a target-tailored source from a small unlabeled operational set. Across toy and real-world targets, it consistently improves over naive source construction and strong holistic/atomistic baselines (Table~\ref{large-table}). Future work will extend the emulator and fingerprint extractor to better capture complex and non-linear pipelines.

\begin{acks}
\footnotesize
This work benefited from access to IDRIS computing resources through the 2025-AD011016555 resource allocation awarded by GENCI and from grant 25-17259K, Fundamental Tradeoffs for Information Hiding in Generated Media (DETERMINE).
\end{acks}

\def\showDOI#1{\unskip}
\def\showURL#1{\unskip}
\bibliography{tada-refs}










\end{document}